\documentclass[aps,rsi,reprint,amsmath,amssymb,floatfix,longbibliography,superscriptaddress]{revtex4-1}

\usepackage{graphicx}
\usepackage{amsmath,amssymb,bbold,bm,color}
\usepackage{siunitx}
\usepackage{float}
\usepackage{bbm}
\usepackage{epstopdf}
\usepackage{hyperref}
\usepackage{booktabs}
\usepackage{comment}
\usepackage[dvipsnames]{xcolor}
\usepackage[normalem]{ulem}

\usepackage[resetlabels,labeled]{multibib}

\newcites{New}{The other list}

\newcommand{\bee}{\begin{equation}}
\newcommand{\ee}{\end{equation}}

\hypersetup{colorlinks=true,linkcolor=blue,citecolor=blue,urlcolor=blue}

\begin{document}

\title{Superconducting diode from flux biased Josephson junction
arrays}

\author{Rafael Haenel}
\affiliation{Department of Physics and Astronomy \& Stewart Blusson Quantum Matter Institute, University of British Columbia, Vancouver BC, Canada V6T 1Z4}
\affiliation{Max Planck Institute for Solid State Research,
70569 Stuttgart, Germany}

\author{Oguzhan Can}
\affiliation{Department of Physics and Astronomy \& Stewart Blusson Quantum Matter Institute, University of British Columbia, Vancouver BC, Canada V6T 1Z4}

\date{\today}
%\pacs{75.30.Ds,62.20.D-,73.43.-f}

%
\begin{abstract}
	We propose a realization of the superconducting diode effect in flux biased superconducting circuits of Josephson junctions. So far the observation of the superconducting diode effect has been limited to 
	rather	
	exotic material platforms. In theoretical proposals, it relied on a non-sinusoidal form of the current phase relation of Josephson junctions.
	Here, we show how the diode effect can be engineered in
superconducting circuits without any such requirements. The only necessary ingredients are standard sinusoidal Josephson junctions and flux bias lines that are readily available in scalable industrial silicon-chip based superconducting design processes.

\end{abstract}

\date{\today}
\maketitle

\section{Introduction}

The superconducting diode effect is present in a  superconducting two-terminal device when it supports a larger critical current in forward direction than in backward direction.
In the ideal limit of maximal imbalance, current applied in one direction is
dissipationless (zero resistance), while it is always dissipative (resistive) in the opposite direction. This
is a natural generalization of the semiconductor diode that is only weakly
resistive in one direction and highly resistive in the opposite direction.

Recent interest in the superconducting diode effect has lead to a surge in
experimental \cite{Wakatsuki2017,Yasuda2019, Ando2020,Pal2021, Lyu2021, Shin2021, Lin2022,Jiang2022,Bauriedl2022, Baumgartner2022,Wu2022}
and theoretical studies
\cite{Hoshino2018, Misaki2021, Kopasov2021, Daido2022, Zinkl2022, Davydova2022, Halterman2022, Scammell2022, Yuan2022, Zhang2022, Zhai2022, Ilic2022, Tanaka2022} including device proposals \cite{Hu2007}. The common factor among them is
reliance on a rather complicated material platform, precluding near-term
commercial applications. The diode effect requires both inversion- and
time-reversal symmetry breaking. Inversion can be absent in
systems with spin-orbit coupling \cite{Ilic2022, Baumgartner2022}, 
or it can be explicitly broken by twisting bilayers \cite{Lin2022, Scammell2022, Ghosh2022}
or applying a current bias \cite{Chiles2022}. 

Two recent theoretical studies proposed a Josephson junction circuit realization of the
superconducting diode effect, independent of an inversion symmetry breaking material platform \cite{Fominov2022, Schrade2022}. However, these
proposals rely on high-harmonic content in the current phase relation of the junctions, 
i.e., a non-sinusoidal current phase relation. Inversion symmetry is
broken by asymmetry of Josephson junctions in the circuit, which differ in their
harmonic content.

In the present work, we propose a realization of the diode effect that
removes all of the above requirements. It is based on a classical circuit of
Josephson junctions with standard sinusoidal current phase relation. Junctions
can in principle even be identical, and inversion symmetry is broken by the
connectivity of the circuit. Time-reversal symmetry is broken by magnetic
fluxes that can be applied using flux bias lines. 
The present diode implementation is agnostic to its underlying material platform and can be
realized in industry standard Nb-Al processes based on scalable semiconductor
technology.

This manuscript is organized as follows. In Sec.~\ref{sec:twojunctions} we
introduce the simplest circuit that gives rise to a diode effect. The approach 
is based on a similar scheme of leveraging high-harmonics in the current phase relation as in \cite{Fominov2022, Schrade2022}, with the key difference that
\textit{effective} high-harmonics are engineered
by chaining identical junctions in series. We generalize this picture in
Sec.~\ref{sec:generalization}, where we give an analytical form for junction
and flux parameters that achieve the most optimal diode for this geometry. In Sec.~\ref{sec:parasites} we discuss effects of parasitic resistors, capacitors, and
inductors.
%and examine the consequences of fabrication variations of Josephson junctions. 
We conclude with a summary in Sec.~\ref{sec:conclusion}.

\section{The diode effect in a superconducting Josephson circuit}
\label{sec:main}
\subsection{Minimal circuit}
\label{sec:twojunctions}

\begin{figure}[ht]
	\centering
	\includegraphics[width=0.6\columnwidth]{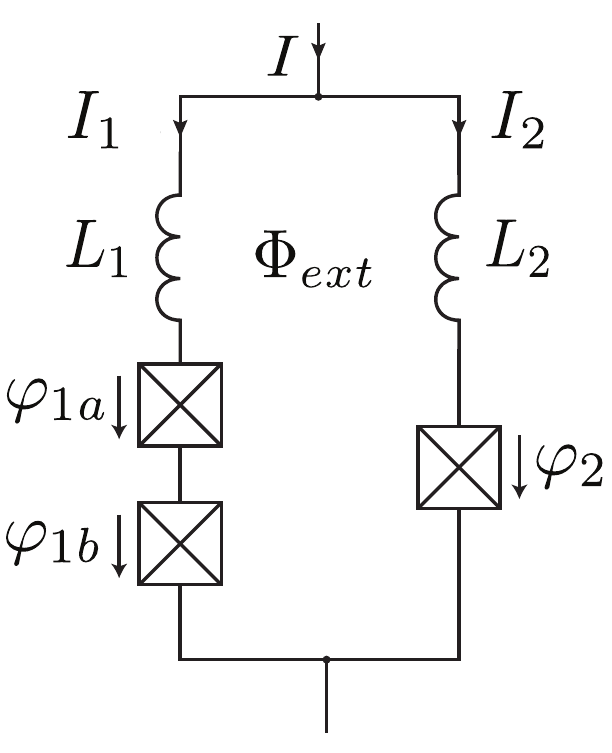}
	\caption{Superconducting circuit consisting of two inductive arms with
	one and two Josephson junctions in series, respectively. An external flux
$\Phi_{ext}$ may thread the superconducting loop. Here, the inductors $L_1,L_2$ represent the geometric inductance of the superconducting wires.}
	\label{fig:circuit1}
\end{figure}

We consider the superconducting circuit depicted in Fig.~\ref{fig:circuit1}
consisting of two arms of Josephson junctions (JJ). The first arm consists of two
junctions across which the superconducting phase jumps by $\varphi_{1a}$,
$\varphi_{1b}$, respectively. The second arm consists of
single Josephson junction with phase difference $\varphi_{2}$. The parasitic inductance of the wires is modeled by the lumped inductors $L_1, L_2$.
This circuit 
has been previously introduced in the context of 
second harmonic
generation, where it is known as the Superconducting Nonlinear Asymmetric Inductive Element
(SNAIL) \cite{Frattini2017}.

The supercurrents in both arms of the SNAIL are determined by the Josephson relations
\begin{align}
    I_1 &= I_c^{(1a)} \sin \varphi_{1a} = I_c^{(1b)} \sin \varphi_{1b} \\
    I_2 &= I_c^{(2)} \sin \varphi_1 \,.
\end{align}

We will be interested in the case of identical junction in the first arm, 
i.e., $I^{(1a)} = I^{(1b)}$. Here, the phases across the junctions are equal,
$\varphi_{1a} = \varphi_{1b} \equiv \varphi_1$.
Continuity of the phase along the superconducting loop then
yields the phase quantization condition
\begin{align}
	2\varphi_1 - \varphi_2 + 2\pi n &= 2\pi \Phi / \Phi_0 \,
\end{align}
where $\Phi$ is the total flux threading the junction and $\Phi_0 = h/2e$ is the
superconducting flux quantum.

For small circuits the geometric
inductance can be neglected, $L_a,L_b \rightarrow 0$.
Then, flux through the loop is solely determined by the
external flux, i.e., $\Phi = \Phi_{ext}$.
The total current is
\begin{align}
	I(\varphi) = I_1 + I_2 = I_c^{(1)} \sin \varphi_1 +  I_c^{(2)} \sin
	\left(2\varphi_1
	-\varphi_{ext} \right)
\end{align}
where we have defined $\varphi_{ext} = 2\pi\Phi_{ext}/\Phi_0$.

The critical current in positive and negative direction are defined as

\begin{align}
	I_c^+ &\equiv \max_\varphi \, I(\varphi) 
	\nonumber\\
	I_c^- &\equiv \left|\min_\varphi \, I(\varphi) \right|
	\label{eq:icm}
\end{align}
where we note that $I_c^+$ and $I_c^-$
are positive quantities.
Importantly, for $\varphi_{ext}/2\pi \notin \mathbf{Z}$, i.e., when $\varphi_{ext}$ is a nontrivial phase,
one generally has
\begin{align}
	I_c^{+}
	\ne
	I_c^{-} \,,
\end{align}
i.e., the forward and backward applied critical currents are imbalanced.
This constitutes the superconducting diode effect. The degree of imbalance between these two critical current values is quantified by the superconducting diode efficiency
\begin{align}
	\eta = \frac{|I_c^{+} - I_c^-|}{I_c^+ + I_c^-} \,,
\end{align}
which is a positive number and $\eta < 1$.
We find the maximum efficiency of $\eta=1/3$ for the two-arm geometry
in Fig.~\ref{fig:circuit1} when $I_c^{(2)} =
I_c^{(1)}/2$ and $\Phi_{ext} = \Phi_0/4$.

\subsection{$N$-arm interferometers and the ideal diode limit}
\label{sec:generalization}
\begin{figure}[t]
	\centering
	\includegraphics[width=\columnwidth]{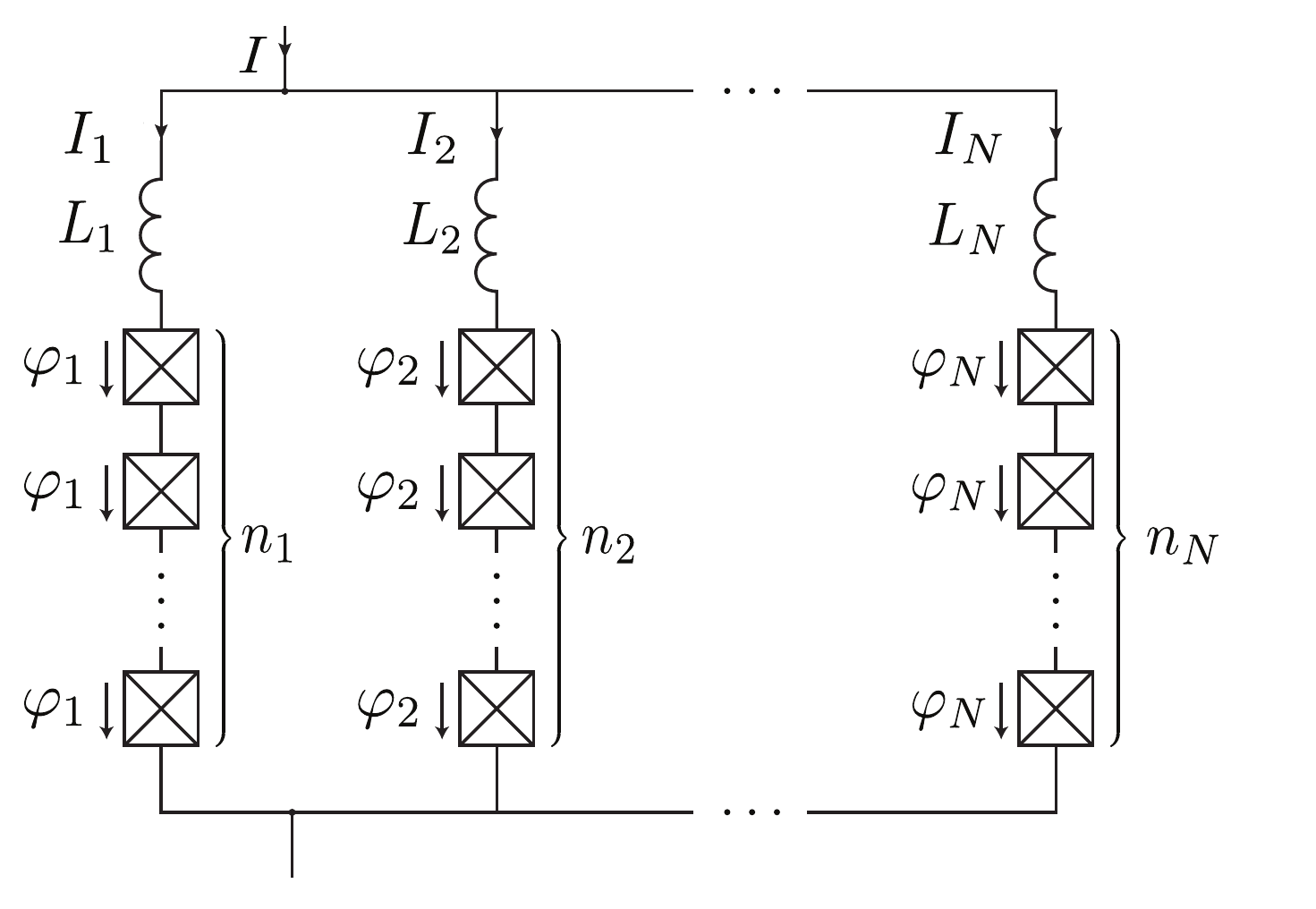}
	\caption{Generalized Josephson interferometer consisting of
	$N$ arms with current $I_k$ and $n_k$ identical
	Josephson junctions with critical current $I_c^{(k)}$ that each experience a superconducting phase drop
	$\varphi_k$. Geometrical inductances have been lumped into 
	the elements $L_k$. A magnetic flux $\Phi_{k,l}$ may be threaded between arms $k$ and $l$.}
	\label{fig:circuit}
\end{figure}
Next we consider a $N$-ladder of JJ arrays which we enumerate by $k=1,..,N$
with $n_k$ consecutive junctions forming each arm $k$ and flux
$\Phi_{k,l}$ enclosed between the arms $k$ and $l$. By phase continuity, the Josephson phase for
a single junction in arm $k$ in terms of the phase variable $\varphi_1$ is given by
\begin{equation}
	\varphi_k = \frac{n_1\varphi_1 -
		\frac{2\pi}{\Phi_0}
	\Phi_{1,k}^{ext} + 2\pi n}{n_k}
	\label{eq:phase-bias-def}
\end{equation} 
in the limit of zero inductance.

We assume the zero-vortex state $n=0$ and define $\phi^{ext}_{k} =
\frac{1}{n_k}\frac{2\pi}{\Phi_0} \Phi_{1,k}^{ext}$.
Then, the total current is
\begin{align}
	I(\varphi_1) = 
	\sum_{k}^{}I_k
	=
	\sum_{k=1}^{N} I_c^{(k)} \sin \left( \frac{n_1}{n_k} \varphi_1 -
	\phi^{ext}_{k} \right) \,.
	\label{eq:igen}
\end{align}
If the integer coefficients $n_k$ are chosen in such a way
that 
\begin{align}
    n_1/n_k=k \,,
    \label{eq:intchoice}
\end{align}
Eq.~\eqref{eq:igen} represents a Fourier series that allows for engineering of arbitrary
current phase relations. 
One such set of integers is given by $n_k = N!/\left[\text{floor}(N/2)!\, k \right]$. For an efficient diode layout, one should however reduce this set by its greatest common denominator, $GCD(\{n_k\})$.

The Fourier series
is fully specified by the
individual magnetic flux parameters
$\phi^{ext}_{k}$ and critical currents $I_c^{(k)}$. 
The fluxes are easily tuned via flux bias lines
and critical currents are determined by the junction area in the fabrication design.

\begin{figure}[t]
	\centering
	\includegraphics[width=\columnwidth]{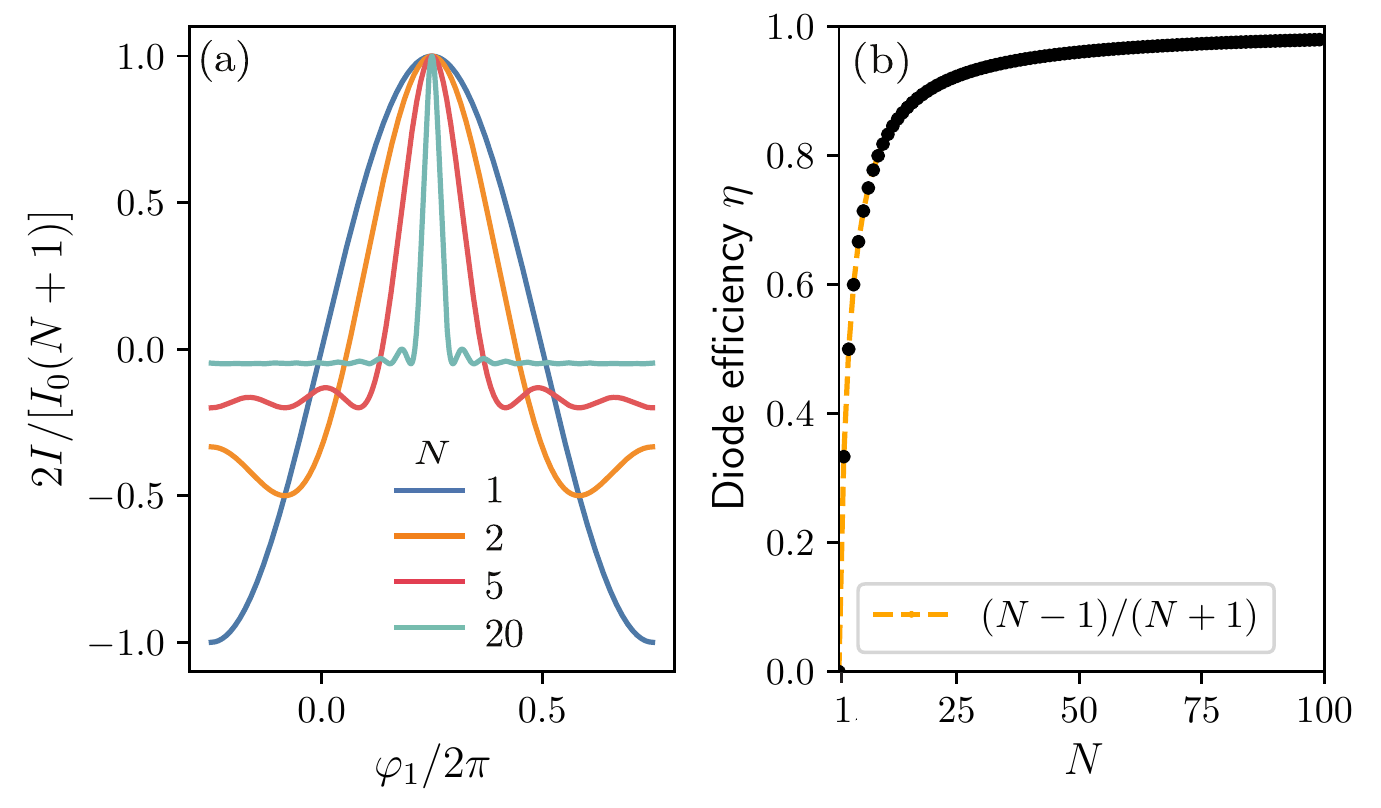}
	\caption{(a) Current phase relation for number of
	interferometer arms $N=1,2,5,20$ and (b) diode efficiency as a
	function of $N$ that is given by $\eta=(N-1)/(N+1)$.}
	\label{fig:efficiency}
\end{figure}

We are now left to discuss the problem of finding the set of
parameters that yield the greatest diode efficiency $\eta$. Let us propose the particular choice
\begin{align}
	\label{eq:param1}
%I_{c}^{(k)} &= I_0 \left( 1-\frac{k}{N+1} \right) \\
	I_{c}^{(k)} &= I_0 \frac{N+1-k}{N} \\
	\phi_{k}^{ext} &= (k-1) \pi/2
	\label{eq:param2}
\end{align}
This yields the following analytical expression for the total current of the
interferometer:
\begin{align}
	I(\varphi_1)/I_0
	&=
	\sum_{k=1}^{N} \frac{N+1-k}{N}\sin \left( k
		\varphi_1 +
		(1-k) \pi/2
	\right)
	\nonumber
	\\
	&=
	\frac{\cos \left((N+1)(\varphi_1-\pi/2) \right)-1}{2N \left(\sin \varphi_1 -1 \right)}
	-\frac{N+1}{2N}
	\label{eq:Ianalytical}
\end{align}
A plot of Eq.~\eqref{eq:Ianalytical} for various $N$ is shown in
Fig.~\ref{fig:efficiency}(a). As $N$ is increased, the function develops a
narrow peak of height $I/I_0=(N+1)/2$ at $\varphi_1 = \pi/2$ over a seemingly flat background
at $I/I_0=(N+1)/2N$, yielding a significant imbalance of critical currents
with diode efficiency
\begin{align}
    \eta = \frac{N-1}{N+1}
\end{align}
that approaches unity in the large-$N$ limit.
In fact, for $N\rightarrow \infty$, the current-phase relation approaches a
series of $\delta$-functions centered at $2\pi (n + 1/4)$ and shifted by a constant current $-I_0/2$, according to
\begin{align}
	\lim_{N\rightarrow \infty}I/I_0 = \pi\sum_k \delta(\varphi_1-\pi/2 + 2\pi k) -
	\frac{1}{2} \,.
\end{align}

While the flux and critical current parameters defined in
Eqs.~(\ref{eq:param1}-\ref{eq:param2}) yield an ideal
diode in the large-$N$ limit, they likely also constitute the fastest converging
series. For small $N\le 5$, where the parameter space is still amenable to numerical optimization, we have numerically confirmed that it represents the optimal solution.

\section{Parasitic components}
\label{sec:parasites}
\subsection{IV-characteristic}
\begin{figure}[t]
	\centering
	\includegraphics[width=\columnwidth]{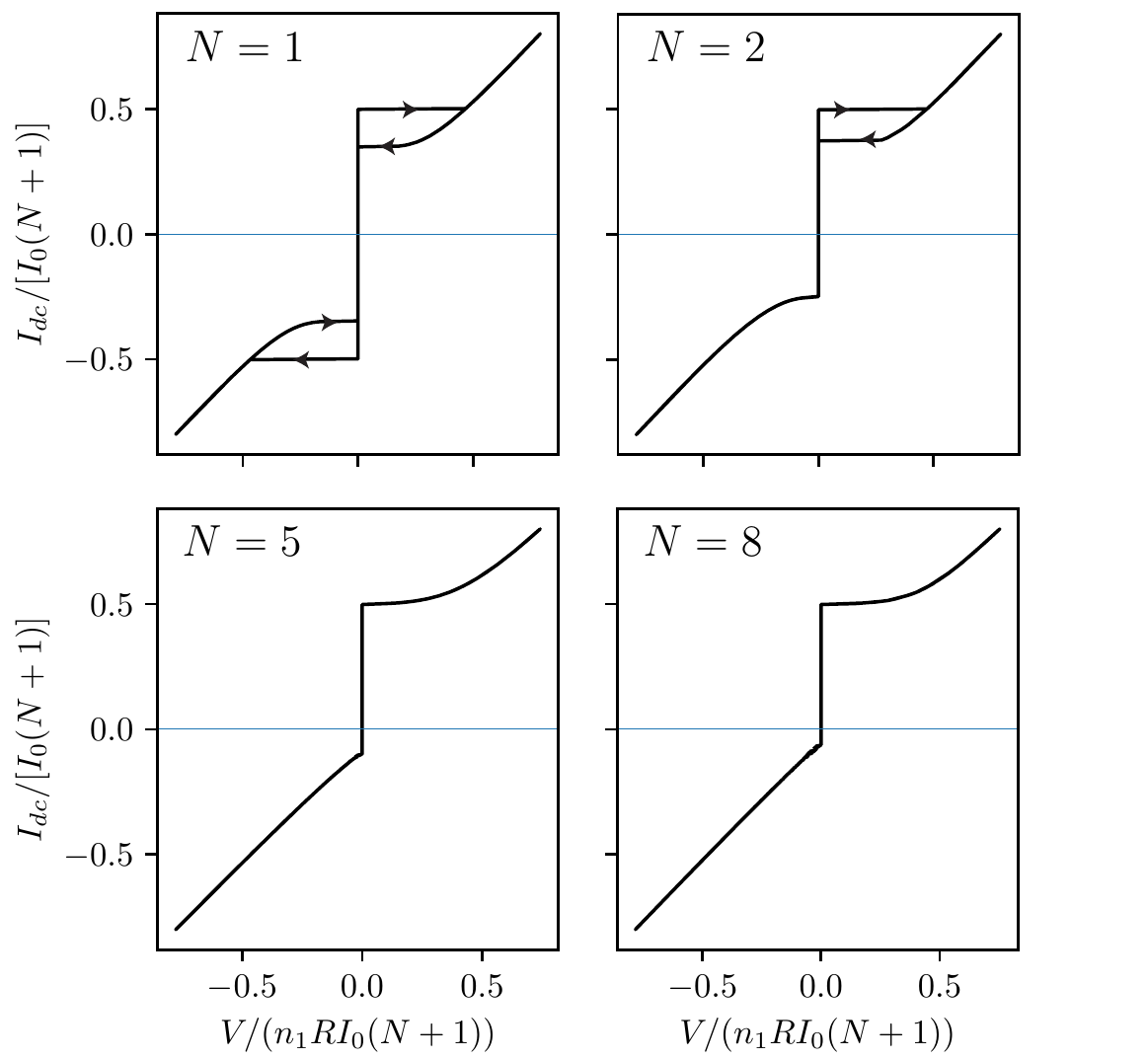}
	\caption{$I$-$V$ curves computed within RSJ model for
	$\beta=0.6\sqrt{n_1}$ and $N=1,2,5,20$}
	\label{fig:rsj}
\end{figure}

We now consider a resistively and capacitively shunted junction (RCSJ) model where each junction on arm $k$ in Fig.~\ref{fig:circuit} is shunted by an internal capacitor $C_k$ and resistor $R_k$. For an external dc current bias $I^{(k)}_{dc}$ in arm $k$ the equation of motion is given by
\begin{equation}
    I^{(k)}_{dc}=\frac{\Phi_0}{2\pi}\frac{\dot \varphi_k}{R_k} + \frac{\Phi_0}{2\pi}C_k\ddot \varphi_k + I^{(k)}_c \sin\varphi_k
\end{equation} 
We assume that every junction is identical for a given arm and current conservation implies that $\varphi_k$ is the same for every junction. Using the phase continuity condition \eqref{eq:phase-bias-def} and adding all currents one can show that the system is governed by a single differential equation
\begin{align}
	I_{dc}/I_0 = \frac{d^2\varphi_1}{d\tau^2} + \beta
	\frac{d\varphi_1}{d\tau} + I(\varphi_1)/I_0  \,,
	\label{eq:diffeq}
\end{align}
where $I_{dc} = \sum_k I_{dc}^{(k)}$ is the total, experimentally applied current bias.
We have also defined the dimensionless parameters
$\tau = \omega_J t$,
$\beta = (\omega_J R_{\text{eff}}C_{\text{eff}})^{-1}$,
$\omega_J = \sqrt{\frac{2e}{\hbar}\frac{I_0}{C_{\text{eff}}}}$ and the effective resistance and capacitance parameters
\begin{align}
    R_{\text{eff}}&=\left(\sum_{k=1}^N\frac{n_1}{R_kn_k}+\frac{n_1}{R_{ext}}\right)^{-1} \\ C_{\text{eff}}&=\sum_{k=1}^N\frac{n_1C_k}{n_k}+n_1C_{ext} \,.
\end{align} 
%which govern the dynamics of the circuit. 
$R_k$ and $C_k$ will depend on the geometries of individual junctions. For better tunability, one may shunt the entire device with an additional resistor $R_{ext}$ and a capacitor $C_{ext}$. The voltage across each arm is given by $V = n_1\frac{\Phi_0}{2\pi}\dot\varphi_1$. In the limit where $R_{ext}\ll R_k$ and $C_{ext} \gg C_k$, dynamics is independent of internal resistances and capacitances of the junctions, where we find $R_{\text{eff}}\approx R_{ext}/{n_1}$ and $C_{\text{eff}}\approx n_1C_{ext}$. In this limit, $\beta$ defined just below \ref{eq:diffeq} scales as $ \beta \approx \sqrt{N!}$
where $\Tilde{\omega} = \sqrt{\frac{2e}{\hbar}\frac{I_0}{C_{ext}}}$ and $n_1$ is the number of junctions on arm $k=1$ which depends on $N$ as $n_k \approx N!$ in the large-$N$ limit

We numerically solve the differential Eq.~\eqref{eq:diffeq} for voltage as a
function of external current $I_{dc}$. The resulting $I$-$V$-curves are shown in
Fig.~\ref{fig:rsj} for various $N$ and $\beta=0.6\sqrt{n_1}$. For $N>1$, the
critical currents become imbalanced. The $I$-$V$-curves display a well-known hysteretic
behavior which, however, can be suppressed for large $N$, i.e., when $\beta$ is sufficiently large.

\subsection{Finite inductance}

\begin{figure}[t]
	\centering
	\includegraphics[width=\columnwidth]{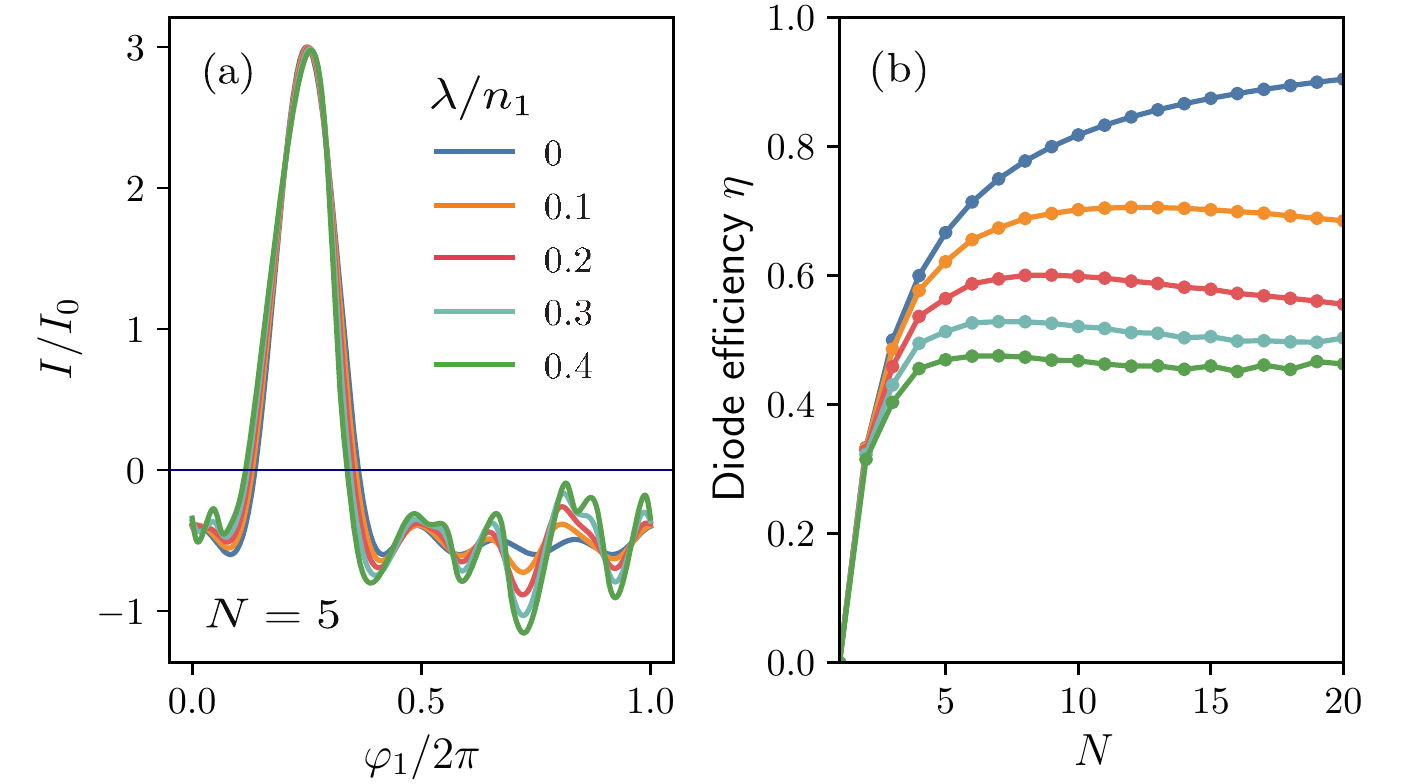}
	\caption{Current phase relation (a) for the exemplary case $N=5$ and (b) diode efficiency as a function of $N$ for varied strength of the inductance parameter $\lambda/n_1$. The geometrical inductance distorts the current-phase relation and diminishes the diode efficiency $\eta$.}
	\label{fig:inductance}
\end{figure}

\begin{table}[t]
\centering
\begin{tabular}{c | c | c | c c c c c c | c c c c c | c c c c c} 
 \hline
 $N$ & $\eta$ & $N_J $ & $n_k$ & & & & & & \multicolumn{4}{l}{$\Phi_{k,k+1}/\Phi_0$} 
 &&\multicolumn{4}{l}{$I_c^{(k)}N/I_0$}
 \\
 \hline
$2$ & $1/3$ & $3$ & 2 & 1 & & & & &$1/4$ & & & & 
&2&1&&
\\
$3$ & $1/2$ & $11$ & 6 & 3 &2 & & & &$3/4$&  $1/4$ & & & 
&3&2&1&\\
$4$ & $3/5$ & $25$ & 12 & 6 & 4 & 3 & && $3/2$&  $1/2$ & $1/4$ & &
&4&3&2&1\\
$5$ & $2/3$ & $137$ & 60 & 30 & 20 & 15 & 12 && $7.5$&  $5/2$ & $5/4$ & $3/4$&
&5&4&3&2&1\\
\end{tabular}
\caption{Optimal parameters for superconducting diode circuits with $N$ arms. $N_J=\sum_k n_k$ is the total number of Josephson junctions required and $\Phi_{k,k+1}$ is the magnetic flux threaded between arms $k$ and $k+1$.}
\label{tab:params}
\end{table}

In the case of finite geometric inductance, $L_k>0$, the flux through the superconducting loops is no longer defined just by the external flux. Instead, Eq.~\eqref{eq:phase-bias-def} must be modified to also include the contribution of the induced flux,
\begin{align}
\varphi_k =
\frac{1}{n_k}\frac{2\pi}{\Phi_0} \left( 
\Phi_{1,k}^{ext} + L_1 I_1 - L_k I_k
\right) \,.
\end{align}
Now, Eq.~\eqref{eq:igen} must be solved self-consistently, since a change in current implies a change in $\varphi_k$, which again induces a change in the $I_k$. For sufficiently large inductances, the induced fluxes will develop hysteretic behavior in the externally applied flux. 

Since the geometrical inductances are expected to be roughly equal for each arm, we parametrize them by a single dimensionless parameter $2\pi L_i/\Phi_0 \equiv \lambda$.
In Fig.~\ref{fig:inductance}, we show the results of the self-consistent calculation for various $\lambda$, small enough that the diode is still in the non-hysteretic regime. The inductance can be seen to distort the current phase relation and yields an overall decrease in diode efficiency. Assuming that the $\lambda$ roughly scales with the maximum number of junctions in a branch $n_1$, this efficiency decrease is more pronounced at large $N$.

%\subsection{Fabrication variations within a single arm}

\section{Discussion and Summary}
\label{sec:conclusion}

We have proposed a superconducting diode circuit element that requires only conventional Josephson junctions and flux bias loops which can be fabricated utilizing existing integrated circuit technology.
%with potential applications in superconducting circuit design. 

A Josephson interferometer will exhibit the superconducting diode effect if current arms carry different harmonics $\sin n \varphi$ of the current phase relation and if these harmonics are phase-shifted with respect to each other.

We have shown that higher harmonics can effectively be generated in generalized SNAIL geometries when multiple conventional Josephson junctions are connected in series. Flux biases in the loops are crucial for generating the phase differences, which we have optimized for diode efficiency. 
Note that polarity of the superconducting diode can be switched by 
reversal of all fluxes, i.e., by setting $\varphi_{k,ext} \rightarrow
-\varphi_{k,ext}$ in Eq.~\eqref{eq:param2}.
We have also considered the effects of geometric inductance, which lead to an overall decrease in diode efficiency.

We work in the limit $E_J \gg E_C $ where charging energy is $E_c = e^2/2C_J$. Here, fluctuations in the phase variable $\varphi$ are suppressed. Therefore $\varphi$ can be treated as a classical variable, justifying the use of RSJ model in Eq.~\eqref{fig:rsj}.

Table~\ref{tab:params} summarizes the various optical circuit parameters for different $N$, along with the resulting diode efficiency in the non-inductive case. Circuits with more than a few arms are likely mostly of academic interest, as the total number of required junctions scales exponentially in $N$, even though more than $10^6$ Josephson junctions and on the order of $10^5$ flux biases have been integrated on a single chip \cite{DWave2010,DWave2021}. Nevertheless, for reasonably small $N$, our proposed diode device could be promising for practical use.

\textit{Note added.} During preparation of this manuscript, we learned about a related experimental work \cite{Gupta2022} that has fabricated and measured the circuit geometry in Fig.~\ref{fig:circuit1} for the simplest case, $N=2$. In their setup, three identical junctions were used, limiting the theoretically achievable efficiency to $\eta=0.28$ instead of $1/3$, which was measured to be $\eta\sim 0.12$ in the experiment.

\section{Acknowledgments}
We thank Marcel Franz, Shannon Egan, and Joe Salfi for enlightening discussions. This work was supported by NSERC, the Max Planck-UBC-UTokyo Centre for Quantum Materials and the Canada First Research Excellence Fund, Quantum Materials and Future Technologies Program.  

\bibliography{scdiode}

\end{document}